\documentclass[a4paper]{jpconf}
\usepackage{graphicx}
\begin{document}
\title{Design, status and test of the Mu2e crystal calorimeter}

\author{N.~Atanov$^1$, V.~Baranov$^1$, J.~Budagov$^1$, R.~Carosi$^5$, F.~Cervelli$^5$, F.~Colao$^2$, M.~Cordelli$^2$, G.~Corradi$^2$, 
E.~Dan\'e$^2$, Y.~I.~Davydov$^1$, S.~Di~Falco$^5$, S.~Donati$^{5,7}$, R.~Donghia$^2$, B.~Echenard$^3$, 
K.~Flood$^3$, S.~Giovannella$^2$, V.~Glagolev$^1$, F.~Grancagnolo$^8$, 
F.~Happacher$^2$, D.~G.~Hitlin$^3$, M.~Martini$^{*,2,4}$, S.~Miscetti$^2$, T.~Miyashita$^3$, 
L.~Morescalchi$^{5,6}$, P.~Murat$^9$, G.~M.~Piacentino$^8$, G.~Pezzullo$^{5,7}$, F.~C.~Porter$^3$,
F.~Raffaelli$^5$, A.~Saputi$^2$, I.~Sarra$^2$, F.~Spinella$^5$,
G.~Tassielli$^{8}$, V.~Tereshchenko$^1$, Z.~Usubov$^1$ and R.~Y.~Zhu$^3$.}

\address{$^1$Joint Institute for Nuclear Research, Dubna, Russia \\
$^2$Laboratori Nazionali di Frascati dell'INFN, Frascati, Italy \\
$^3$California Institute of Technology, Pasadena, United States \\
$^4$Universit\`a Guglielmo Marconi, Roma, Italy \\
$^5$INFN Sezione di Pisa, Pisa, Italy \\
$^6$Dipartimento di Fisica dell'Universit\`a di Siena, Siena, Italy \\
$^7$Dipartimento di Fisica dell'Universit\`a di Pisa, Pisa, Italy \\
$^8$INFN, Sezione di Lecce and Dipartimento di Matematica e Fisica dell'Universit\`a del Salento, Lecce, Italy. \\
$^9$Fermi National Laboratory, Batavia, Illinois, USA}

\ead{matteo.martini@lnf.infn.it}

\begin{abstract}
The Mu2e experiment at Fermilab searches for the charged-lepton flavor violating neutrino-less conversion of a negative 
muon into an electron in the field of a aluminum nucleus. The dynamic of such a process is well modeled by a two-body
decay, resulting in a monoenergetic electron with an energy slightly below the muon rest mass ($104.967$ MeV). 
The calorimeter of this experiment plays an important role to provide excellent particle identification capabilities
and an online trigger filter while aiding the track reconstruction capabilities. The baseline calorimeter configuration 
consists of two disks each made with $\sim 700$ undoped CsI crystals read out by two large area UV-extended 
Silicon Photomultipliers. These crystals match the requirements for stability of response, high resolution 
and radiation hardness. In this paper we present the final calorimeter design.
\end{abstract}

\section{Introduction}
The Mu2e experiment at Fermilab~\cite{art1} aims to search for Charged Lepton Flavor Violation (CLFV) in the neutrino-less, coherent conversion of a 
negative muon into an electron in the Coulomb field of an $^{27}$Al nucleus. The $\mu\to e$ conversion results in a mono energetic electron 
with an energy equal to the muon rest mass minus the corrections for the nuclear recoil and the binding energy of the muon. For $^{27}$Al
the energy of the mono energetic electron is equal to $104.97$ MeV. 

The experiment is designed to reach the single event sensitivity of $2.4\times 10^{-17}$ in three years  \cite{art1}. This value represents an improvement 
of four order of magnitude over the current best experimental limit set by SINDRUM II experiment \cite{art2}.

The Standard Model predicted rate for this process is $\mathcal{O}(10^{-52})$~\cite{art3}, therefore any signal observed by Mu2e would be a compelling evidence of new physics.

\section{Mu2e electromagnetic calorimeter}

The Mu2e detector is designed to be almost background free and it is located inside a large superconducting solenoid with a magnetic field 
of 1 T in the region of the calorimeter. This detector is located just behind the tracker and complements it by providing: powerful $\mu/e$ particle 
identification, seed for pattern recognition in the tracker and an independent software trigger system. Efficient particle identification requires a time resolution better 
than 500 ps and an energy resolution $\mathcal{O}(5\%)$.

The calorimeter should be able to operate in an environment where a dose up to 100 krad and a neutron fluency of $10^{12}$ n/cm$^2$ are expected. It must 
also works in a 1 T magnetic field and $10^{-4}$ Torr vacuum ensuring the redundancy of the component. 

\begin{figure}[h]
\begin{center}
\begin{minipage}{14pc}
\includegraphics[width=12pc]{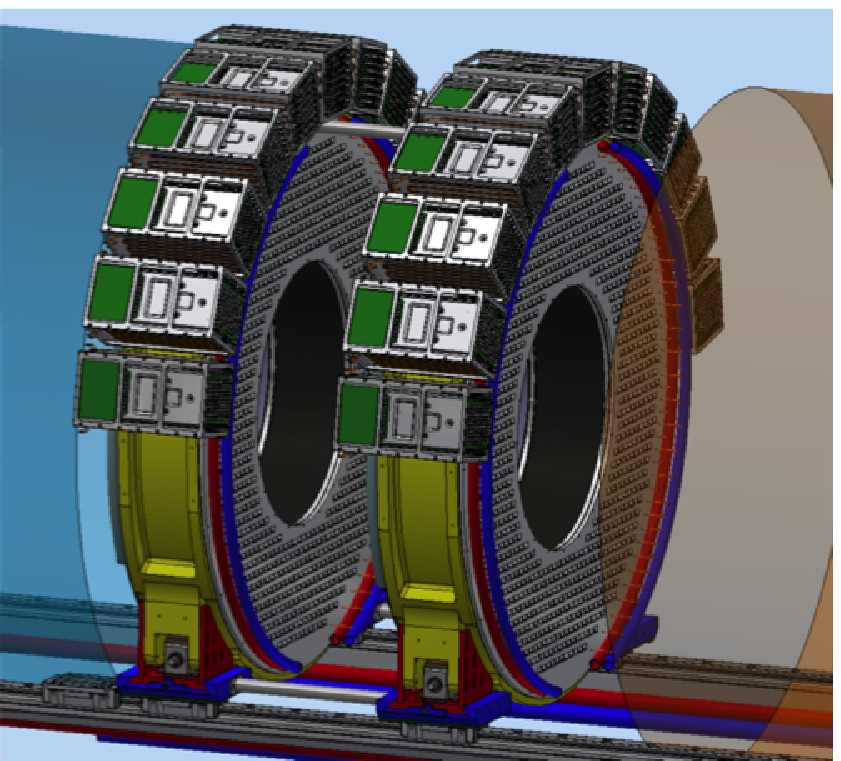}
\caption{\label{figura1}Structure of the Mu2e calorimeter}
\end{minipage}\hspace{2pc}%
\begin{minipage}{14pc}
\includegraphics[width=16pc]{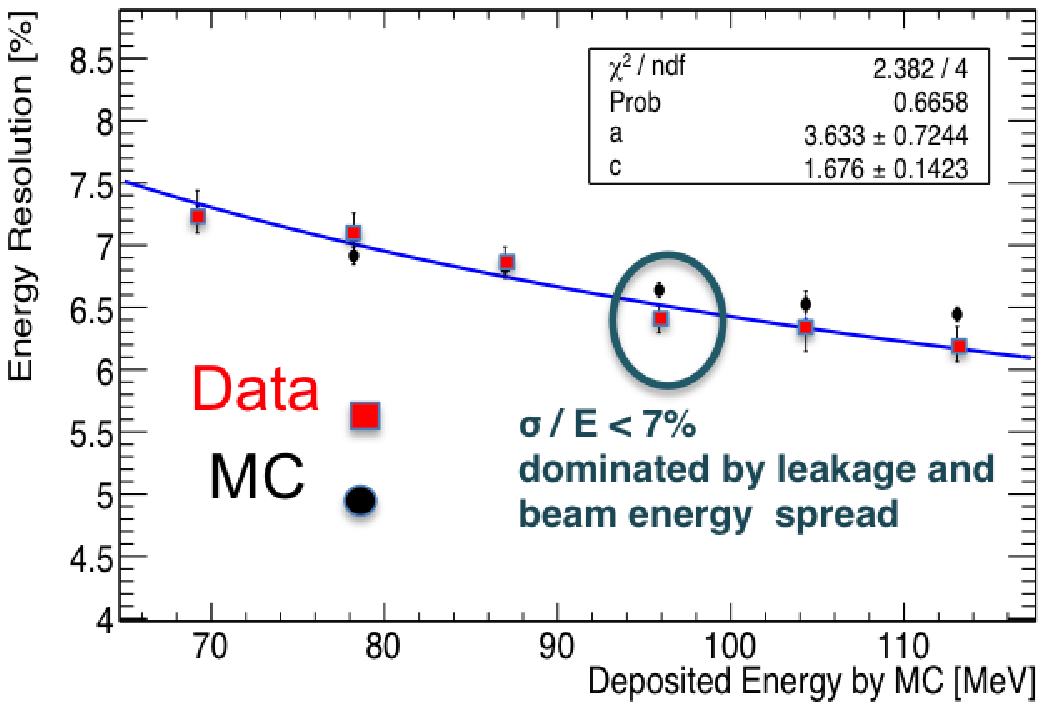}
\caption{\label{figura2}Measured energy resolution for a $3\times 3$ CsI matrix with Hamamatsu TSV $12\times 12$ mm$^2$ MPPCs using electrons between 80 and 120 MeV.}
\end{minipage} 
\end{center}
\end{figure}

Due to physical and geometrical constraints, we decided to adopt a solution with two annuli made by undoped CsI crystals each read out using two 
Silicon Photomultipliers. Each disk has an internal (external) radius of 374 mm (660 mm) and is filled with $34\times 34\times 200$ mm$^3$ crystals (see Fig.\ref{figura1}). The disks 
are separated by about half electron wavelength (75 cm). The analog read out electronics is connected to the SiPM while the digital boards are housed in the crates in the top of each 
disk. 

The discussed requirements pushed the experiment to adopt a calorimeter made by undoped CsI crystals optically coupled to $14\times 20$ mm$^2$ large area UV-extended SiPM. 

\section{Crystal choice and test}

The requirements of the electromagnetic calorimeter imply to use crystals with:
\begin{itemize}
\item high light output
\item good light response uniformity, LRU, ($>10\%$)
\item fast signal with small component ($\tau<40$ ns)
\item radiation hard with maximum light output loss below 40\%
\item small radiation induced readout noise (below 0.6\%)
\end{itemize}

In the Conceptual Design Report~\cite{art4}, the baseline calorimeter choice was LYSO crystals readout with APD and many 
tests were carried out for this option~\cite{art5}. A large increase on the Lu$_2$O$_3$  salt price in 2013 made this option 
unaffordable, so that for the TDR~\cite{art1} we have opted for cheaper crystals such as BaF2 and CsI. After a long R\&D 
program~\cite{art6}, we have finally selected undoped CsI crystals as baseline choice (See Tab.\ref{tabella1}).

\begin{center}
\begin{table}[h]
\centering
\caption{\label{tabella1}Crystals suitable for Mu2e calorimeter. } 
\begin{tabular}{@{}l*{15}{l}}
\br
Crystal & $BaF_2$ & LYSO & CsI & $PbWO_4$ \\
\mr
Density [$g/cm^3$] & 4.89 & 7.28 & 4.51 & 8.28 \\
Radiation Length [cm] & 2.03 & 1.14 & 1.86 & 0.9 \\
Moliere radius [cm] & 3.10 & 2.07 & 3.57 & 2.0 \\
dE/dx [MeV/cm] & 6.5 & 10.0 & 5.56 & 13.0 \\
Refractive Index at $\lambda_{MAX}$ & 1.50 & 1.82 & 1.95 & 2.20 \\
Peak Luminescence [nm] & 220/300 & 402 & 310 & 420 \\
Decay time [ns] & 0.9/650 & 40 & 26 & 30/10 \\
Light Yield [\% NaI] & 4.1/3.6 & 85 & 3.6 & 0.3/0.1 \\
Light Yield Variation with temperature [\%/$^oC$] & 0.1/-1.9 & -0.2 & -1.4 & -2.5 \\
Hygroscopicity & None & None & Slight & None \\
\br
\end{tabular}
\end{table}
\end{center}

Tests on CsI crystals have been performed for three different vendors: ISMA (Ukraine), SICCAS (China) and Optomaterial (Italy). 
These crystals have been irradiated 
up to 900 Gy and to a neutron fluency up to $9\times 10^{11}$ n$_{1MeV}$/cm$^2$. The ionization dose does not modify LRU while a 20\% 
reduction in light yield has been observed at 900 Gy. Similarly, the neutron flux causes a 15\% light yield deterioration. These results 
are compatible with the requirements of the calorimeter. 

A small unroped CsI matrix has been built and tested in Frascati Beam Test Facility using electrons with energy between 80 and 120 MeV. The prototype is 
a $3\times 3$ matrix and each crystal is read out using an array of sixteen $3\times 3$ mm$^2$ Hamamatsu TSV MPPCs. During this test we measured 
a light yield of 30 (20) pe/MeV with (without) optical grease with Tyvek wrapping. The measured time and energy resolution, 110 ps and 7\% respectively, 
perfectly match our initial requirements (Ref. Fig.\ref{figura2}).

\section{UV-extended SiPM}

The requirement of having a small air gap between crystal and photodetector and the request of redundancy in the read out implies the use 
of custom SiPMs. For the Mu2e experiment we have increased the transversal dimension of the CsI from $30\times 30$ to $34\times 34$ mm$^2$ in order to 
accomodate two $2\times 3$-arrays of $6\times 6$ mm$^2$ UV-extended SiPM. This allows to work with an air-gap while satisfying the pe/MeV requirement 
with a single SiPM. Presently, we are testing both Hamamatsu MPPCs and FBK UV-extended Silicon Photomultipliers. 

The photosensors are packaged using a parallel arrangement of two groups of three cells biased in series. The  
samples already acquired show a good PDE ($\sim 30\%$ at 315 nm) with a gain greater than $10^6$ at an over-voltage of 3 V with respect to the breakdown voltage.
The series connection produce a signal with a total width of 70 ns. A first array has been assembled with 6 Hamamatsu 6x6 MPPCs and has 
been optically connected to a CsI Tyvek wrapped crystal measuring a time 
resolution for 1 MIP ($\sim 20$ MeV) of 170 ps. 
 
Following calorimeter requirements, one important aspect to be considered for the read out of the crystals is the radiation hardness of the SiPM. In this 
context, we have performed different tests using Hamamatsu and FBK UV-extended devices. SiPM irradiated with  a dose of up to 20 krad photons source do not show 
any effect on the leakage current. Different effects are observed with neutrons. When exposing sensors to 14 MeV neutrons with a total flux of $2.2\times 10^{11}$ n/cm$^2$ 
(corresponding to 2.2 times the 
experiment lifetime), we observe a too high increase of the leakage current (up to 2.3 mA when using $3\times 3$ mm$^2$ Hamamatsu cells). 
To reduce the leakage current to acceptable value, we need to cool down all SiPM to a temperature of 0 degrees. In order to do so, we will use a dedicated cooling station for the calorimeter. 
The final choice of the coolant and the parameters of the station is currently under study.

\begin{figure}[h]
\begin{center}
\begin{minipage}{14pc}
\includegraphics[width=14pc]{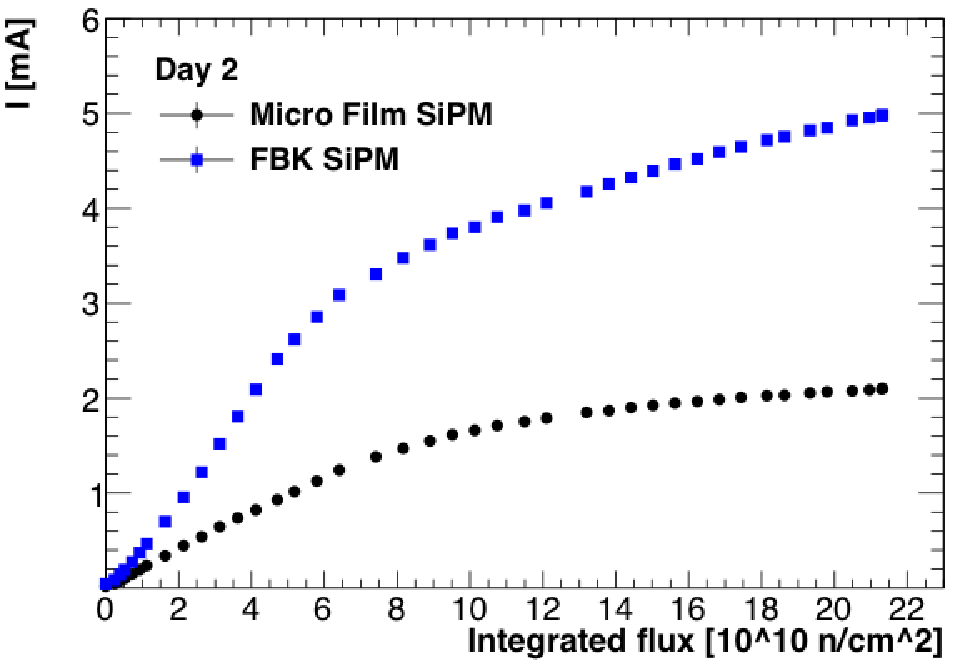}
\caption{\label{figura3}Measured leakage current versus integrated neutron flux.}
\end{minipage}\hspace{2pc}%
\begin{minipage}{14pc}
\includegraphics[width=14pc]{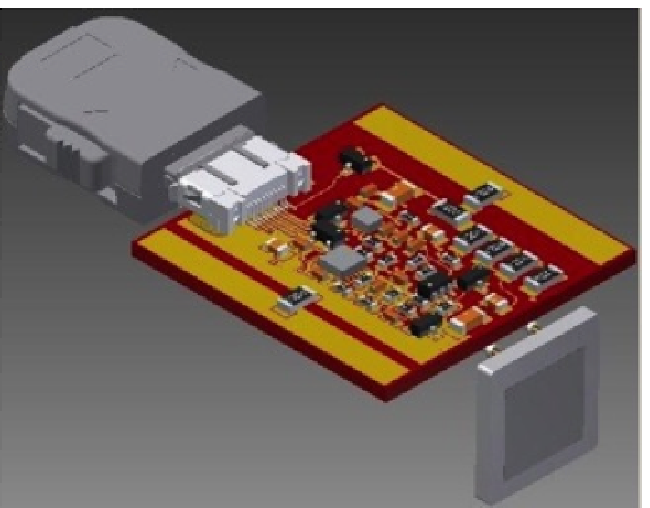}
\caption{\label{figura4}Rendering of the FEE board connected with SiPM. }
\end{minipage}
\end{center} 
\end{figure}

\section{Front End Electronics}

Each SiPM is directly connected to a dedicated board (See Fig.\ref{figura4}) housing a transimpedence preamplifier with a settable gain $\times 15$ or $\times 30$, 2~V 
dynamic range and 15 ns rise time. This board provides also a pulse signal for testing the preamplifier and a slow control readout of temperature and leakage current. 
The digital 
boards are housed into 11 crates per disk with 20 differential channels per board. These boards host a mezzanine which: receives signals from SIPM, manage HV setting 
and includes a Waveform Digitizer section that is based on  SmartFusion II FPGA with 200 Msps 12 bit ADC. 

%\section{Absolute calibration and system monitoring}
%The absolute energy scale, the equalization and the linearity of the response are provided by a fluorinert source system which is pumped through 
%pipes that covers the entire disk front face surface. This liquid can be activated by a neutron source thus producing 6.13 MeV photons. The designed 
%piping scheme allows to get a uniform illumination of the disk with variation on the intensity less than 5\%. 
%The variation of the crystal optical transmittance and photosensors gains will be monitored in a continuos way by means of a laser system, following a scheme 
%similar to the one used for CMS experiment\cite{art7}. Prototype of the distribution system has been developed and tested.  

\section*{Acknowledgments}

This work was supported by the EU Horizon 2020 Research and Innovation
Programme under the Marie Sklodowska-Curie Grant Agreement No. 690835.

\end{document}